\documentclass[number, 3p]{elsarticle}
\usepackage[english]{babel}
\usepackage[utf8]{inputenc}
\usepackage{amsmath}
\usepackage{graphicx}

\usepackage[normalem]{ulem}
\usepackage{soul}

\usepackage[usenames,dvipsnames,svgnames,table]{xcolor}

\usepackage{color}

\begin{document}
\title{Economic Inequality between Groups in an \emph{a priori} Stratified Society}
\author[utfpr]{Thiago Dias}
\ead{thiagodias@utfpr.edu.br}
\author[ufrgs]{Sebastián Gonçalves}
\ead{sgonc@if.ufrgs.br}

\address[utfpr]{Universidade Tecnológica Federal do Paraná, Campus Dois Vizinhos, Estrada para Boa Esperança, km 04, 85660-000, Dois Vizinhos PR, Brazil}
\address[ufrgs]{Instituto de Física, Universidade Federal do Rio Grande do Sul, Caixa Postal 15051, 91501-970 Porto Alegre RS, Brazil}
\date{\today}
\cortext[ca]{Corresponding author}

\begin{abstract}
We present an agent-based model of economic exchange in a society composed of two groups, representing two social groups and with different internal protection rules for the poor agents. The goal is to address the emerging wealth distribution when economic rules are not the same for all individuals. Individuals exchange wealth in pairwise interactions with no underlying lattice. The wealth, risk aversion factor, and group of the agents characterize their state. The wealth exchanged between two agents obeys a fair rule: the quantities put at stake by them are the same regardless of who wins. One agent can interact with another agent in the same or the other group, controlled by a rate which is a parameter of the model.  Inter-group exchanges obey an exclusive protection rule, which can be understood as a public policy to reduce inequality. We show that the most protected group accumulates more wealth, has less inequality, and has higher mobility than the other group. The results of simulations are compared with income distribution in Brazil discriminated by race as an example of the application our model.
\end{abstract}


\maketitle

\section{Introduction}
When analyzing the wealth or income of different economies, one can notice similarities. A ubiquitous feature is the division of societies into two classes. A minority wealthy class, described by a Pareto power-law for the income distribution~\cite{pareto, boghosian}, and a majority with incomes distributed in a way that can be fitted by a Gibbs distribution or log-normal functions~\cite{dragulescu}. Most of the time, these results are obtained by considering only the income or wealth of individuals.

Nevertheless, societies have other forms of distinguishing their individuals.  Inequalities among ethnicity, gender, and race are examples of divided societies where people belonging to a particular subgroup have status, power, and wealth~\cite{berreman,anthias, grusky}. Indeed, in many societies, dominant classes are composed of people associated with one of the groups, and economic inequality might be related to this non-economical classification~\cite{campante,kilsztajn, shaikh, darity}.

Brazil is one example of how social stratification by race and gender affects economic inequality. According to the Brazilian Institute of Geography and Statistics (IBGE-2020)~\cite{ibge}, approximately 43\% of the country population are White, and 56\% consider themselves Afro-Brazilian (the racial group composed of Black and Brown individuals). The other 1\% is formed by Yellow and Indigenous. We want to point out that, due to miscegenation, the race data is from self-declaration. From 2018 to 2020, the average income of Whites remained  50\% larger than the average income of Afro-Brazilian. The disparity is even more considerable between White male and Black/Brown female incomes, the difference exceeding 150\%.

The field of econophysics has already proven to be very valuable with first principles and microscopic models of pairwise wealth exchange capable of reproducing important economic actual data~\cite{yakovenko, burda, cardoso}. These models consider ensembles of economic agents that interact, exchanging part of their wealth. Wealt h, in this case, can be understood as money given for some commodity or service \cite{iglesias}. 

Although simulations and analytical results using the econophysics approach can reproduce Gibbs distribution and Pareto tails in the so-called ``two-class'' regime, to our knowledge, no contribution makes a distinction among individuals besides the economic features.

We study the economy of a society composed of two social groups, where individuals are characterized by their wealth, saving tendencies, and the group to which they belong. The transactions inter and intra-group obey different rules. In the next section, we describe the model in detail.

\section{Kinetic exchange model with two social groups}
The dynamics of the economies run through agent-based simulations in which agents belong to one of two groups. Differently from work done by Lim and Min \cite{lim}, the groups here are not related to wealth; they represent different races, gender, or castes. We characterize the agents by their wealth $w_i$, risk-aversion factor $\beta_i$, and the group they belong to, labeled $A$ and $B$. The risk-aversion factor sets the fraction of wealth that an agent puts at stake during a transaction, which is $(1-\beta_i)w_i$ \cite{iglesias-12}.

At each Monte-Carlo step (MCS), pairs of agents are chosen, so every individual has one exchange on average. Taking agents $i$ and $j$ to make an exchange and assuming the former wins, we have
\begin{align}
    w_i(t+1) &= w_i(t) + \Delta w\nonumber\\
    w_j(t+1) &= w_j(t) - \Delta w,\label{exchange}
\end{align}
where $w_{i(j)}(t+1)$ represents the wealth of agent $i(j)$ after, and $w_{i(j)}(t)$ before the transaction. Different ways of determining $\Delta w$ can be found in the literature \cite{iglesias-12, caon, hayes}. To mimic actual wealth exchanges, the amount put at stake must be the same by the two individuals, which we guarantee using 
\begin{equation}
\Delta w = \min[(1-\beta_i)w_i(t), (1-\beta_j)w_j(t)].
\end{equation}

It is important to stress that while this represents theoretically ``fair'' exchanges, it leads the system to the condensate state \cite{cardoso-21}, where one or few agents have everything and the rest nothing. Condensation seems to be the path the global economy follows, since the inequality grows continuously \cite{piketty}. 

Some modifications have been proposed to overcome condensation. Here we study the impact of the social protection that increases the probability of the poorest agent to win the trade, first proposed by Scafetta \cite{cardoso, caon, scafetta}, which depends on the wealth of the agents $i$ and $j$ prior to the exchange:
\begin{equation}
    p = \frac{1}{2} + f\frac{|w_i(t) - w_j(t)|}{w_i(t) + w_j(t)}.
    \label{protection}
\end{equation}

In equation~\ref{protection}, $f$ represents a social protection factor which can be a governmental regulation or, as in the present case, different rules regarding trades between agents of the same group as compared with the other group. The protection factor can assume values from 0 to 0.5. The value 0 (no protection) corresponds to equal probabilities (0.5) of winning for both agents, while the value 0.5 (highest protection) always favors the poor agent. The richest agent has the complementary probability $1-p$ of win.

Interactions between agents belonging to the same group have their own social protection factors. For example, when two agents of group $A(B)$ exchange wealth, $f_{A(B)}$ increases the probability of the poorer agent wins. Inter-group trades respect the protection given by the $f$, for $A\rightarrow B$ and $B\rightarrow A$ transactions. Once $f$ corresponds to a public policy to reduce inequality, in real economies the minimum value of $f_{A}$ and $f_B$ must be $f$.


We measure inequality of each group by using the Gini index, proposed by C. Gini in 1912, and calculated by
\begin{equation}
    G_{I}(t) = \frac{1}{2W_I(t)N_I}\sum_{ij}^{N_I} | w_i(t) - w_j(t)|,
    \label{gini}
\end{equation}
where $I = A,B$ indicates the group over which the index is calculated and $W_I = \sum_i^{N_I} w_i$ the wealth of group $I$. $G$ for the whole society can be calculated by doing the summations over all agents. The Gini index varies from 0 (perfect equality) to 1, which corresponds to the condensate state, where only one individual possesses all the available wealth \cite{firebaugh}.

The liquidity $L$ is the amount of wealth that is being exchanged at each time step normalized by the total group wealth. Given that the total wealth is conserved, enrichment and impoverishment are results solely of the exchanges. Consequently, liquidity is a measure of the economic mobility. Larger values of $L$ indicate that more agents are exchanging their wealth whereas $L\rightarrow 0$ when the economy gets stagnated and condenses. We calculate liquidity \cite{iglesias-12} as
\begin{equation}
    L_I(t) = \frac{1}{2W_I(t)}\sum_i^{N_I} |w_i(t) - w_i(t-1)|.
    \label{liq}
\end{equation}

The simulations are performed with a system of 1000 individuals --equally divided into groups A and B--, and we make averages over 200 ensembles of $5\times10^4$\,MCS each one for each set of key parameters. As initial conditions, $w_i$ and $\beta_i$ are uniformly distributed in the interval (0,1); while $\beta_i$'s are fixed, $w_i$'s are functions of time due to the the agents' exchanges. We avoid very low $\Delta w$ values by choosing a minimum wealth an agent must held to participate of the exchanges, $w_{min} = 1\times10^{-9}$. Thus, agents with $w_i < w_{min}$ are excluded from the economy. 

During the first $10^3$\,MCS only trades among members of the same group were allowed. Then the two groups are put in contact (much like two gases in a mixture) and inter-group exchanges can take place with probability $p_{AB}$, which is fixed during the simulations. Intra-group transactions occur with probability $1-p_{AB}$. 
The key parameters we analyze in this work are the three social protection factors ($f_{A}$, $f_B$, and $f$) and the probability of inter-group trades ($p_{AB}$).

\section{Results}

\begin{figure*}[!htbp]
    \centering
    \includegraphics[width = 0.9\textwidth]{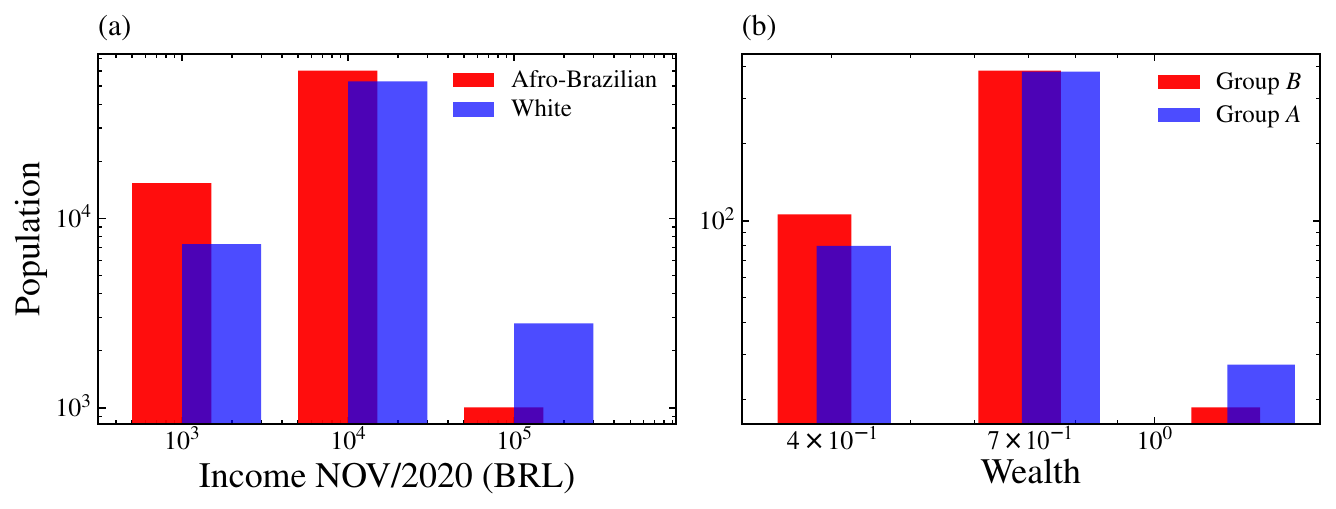}    
    \caption{Qualitative comparison between (a) income distribution for Afro-Brazilian and White individuals in November of 2020 and (b) model numerical simulations with $f_A = 0.4$, $f = f_B = 0.33$, and $p_{AB} = 0.5$ in log-log scale. Data in the left panel was collected from~\cite{pnad}. Blue bars are shifted horizontally for better visualization. Please notice the different horizontal and vertical scales. }
    \label{fig:compare}
\end{figure*}

For illustration purpose, Figure~\ref{fig:compare} shows a qualitative comparison of income distributions for Afro-Brazilians and Whites in November 2020~\cite{pnad} with the results of our model for $f = f_B = 0.33$, $f_A = $0.40 and equal probabilities of inter and intra group exchanges.
It is noticeable in the left panel that most individuals in the lowest income class are Black and Brown, while the opposite is true for those with the highest income.

The work of Campante \emph{et al.} \cite{campante} revealed that not only educational level and insertion in labor market are parameters to define the income of a person. Using an adapted Oaxaca-Blinder methodology \cite{oaxaca, blinder, bourguignon}, they argued that there is a racial discrimination parameter that plays a role in the determination of wage of Black and Brown individuals in Brazil.

Nobel prize winner D. Card and his collaborators argued that non White workers are more likely to be sorted in lower-wage positions in Brazil.
Due to the educational gap between Whites and Afro-Brazilians, one can expect more White workers in higher-level positions. Yet it represents only half of the contribution to the racial payment divergence. The remaining ``unexplained'' 50\% may be related to discriminatory policy~\cite{gerard}. We include asymmetry in the present model, and therefore a difference for total wealth and Gini index, by means of the social protection factor, i.e. making $f_A \neq f_B$.

\begin{figure}[!htbp]
    \centering
    \includegraphics[width = 0.9\textwidth]{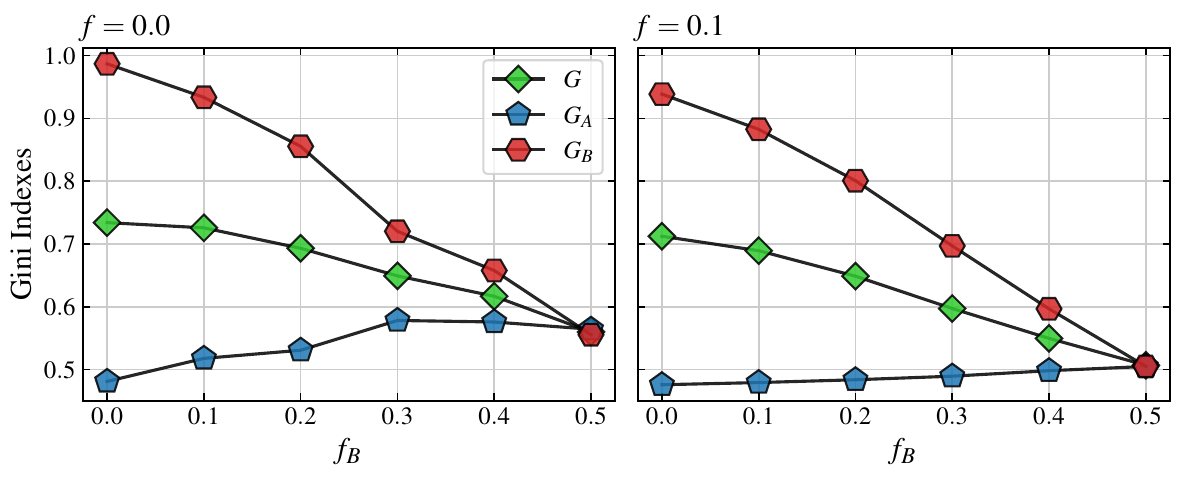}
    \caption{Gini indexes $G_A$ and $G_B$ of groups $A$ and $B$, respectively, and $G$ for the whole system as functions of the social protection factor $f_B$ of group $B$ and for $f = 0.0$ (left) and 0.1 (right). Simulations were carried out with $f_A = 0.5$ and $p_{AB} = 0.1$. Lines are guides to the eyes.}
    \label{fig:f2g}
\end{figure}

We start the simulations with $f_A = 0.5$ and $p_{AB} = 0.1$ fixed, and $f_B$ was our free parameter. Figure~\ref{fig:f2g} shows Gini indexes for both subsystems ($G_A$ and $G_B$) and the whole economy ($G$) as functions of $f_B$. We considered two values of $f$: 0.0 and 0.1. $G$ and $G_B$ reduce with increasing $f_B$ in both cases. On the other way, $G_A$ gets larger with $f_B$. The inter-group protection also affects the inequality, larger $f$ implies in smaller $G$, $G_A$, and $G_B$ for all $f_B$ we analyzed. Even resulting in an increase of the inequality of group $A$, the protection of $B$ results in a more egalitarian society as a whole.

\begin{figure}[!htbp]
    \centering
    \includegraphics[width = 0.5\textwidth]{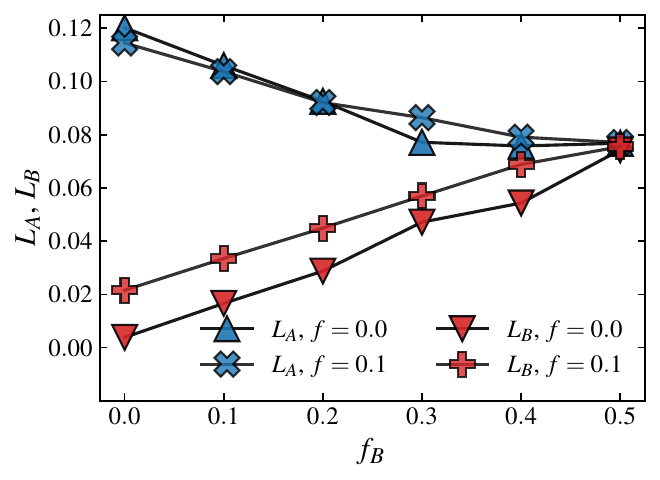}
    \caption{Liquidities $L_A$ and $L_B$ of the groups $A$ and $B$, respectively, as functions of $f_B$. The fixed parameters in the simulations were$f_A = 0.5$ and $p_{AB} = 0.1$. Two inter-group social protection factors were considered, $f = 0$ and $0.1$.} 
    \label{fig:f3l}
\end{figure}

Wealth-conservative systems present negative correlations between Gini and economic mobility. Lower inequality corresponds to larger mobility and vice-versa.\cite{bertotti,nener} Our results corroborate these findings. We notice a boost in the liquidity of $B$ as a result of growing $f_B$ (which also diminishes $G_B$). In figure~\ref{fig:f3l} we show $L_A$ and $L_B$ as functions of $f_B$. Two inter-group protection factors, $f = 0.0$ and 0.1, were considered, and $p_{AB} = 0.1$ remained fixed. As one can see $L_A$ is also affected by $f_B$, but in a negative way. We have shown in figure~\ref{fig:f2g} that $f_B$ increases the inequality of group $A$. Thus, a decrease of $L_A$ with $f_B$ should be expected. It is undesirable to the agents of $A$, since their mobility is reduced, but this is beneficial to the entire system's economy.

\begin{figure}[!htbp]
    \centering
    \includegraphics[width = 0.5\textwidth]{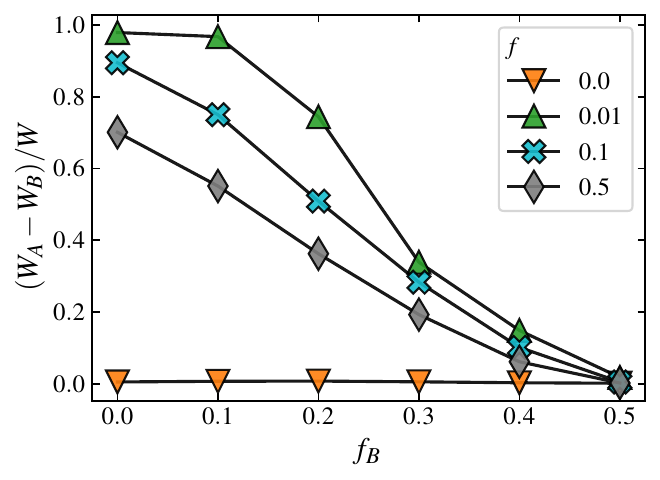}
    \caption{Wealth transferred from $B$ to $A$ normalized by the total available wealth for different as function of $f_B$ for $f = 0$, 0.01, 0.1, and 0.5. $p_{AB} = 0.1$ and $f_A = 0.5$ were kept fixed during the simulations. Lines are guides to the eyes.} 
    \label{fig:f4w}
\end{figure}

We show the transfer of wealth from $B$ to $A$ in figure~\ref{fig:f4w}, where we observed that the wealth transferred is smaller as $f_B$ gets closer to 0.5 and disappears when $f_A = f_B = 0.5$. Inter-group social protection also interferes on how much wealth flows from one group to the other. Four different $f$ were considered: 0.0, 0.01, 0.1, and 0.5. Surprisingly, for unprotected intergroup exchanges, no net wealth is transferred between the groups. It seems that $f > 0$ promotes the accumulation of wealth by the most protected group if $f_A \neq f_B$. The case of $f = 0.01$, a very small inter-group protection, causes an accumulation of more than 95\% of the available wealth by agents of $A$ when $f_B = 0$ and 0.1.

\begin{figure}[!htbp]
    \centering
    \includegraphics[width = 0.5\textwidth]{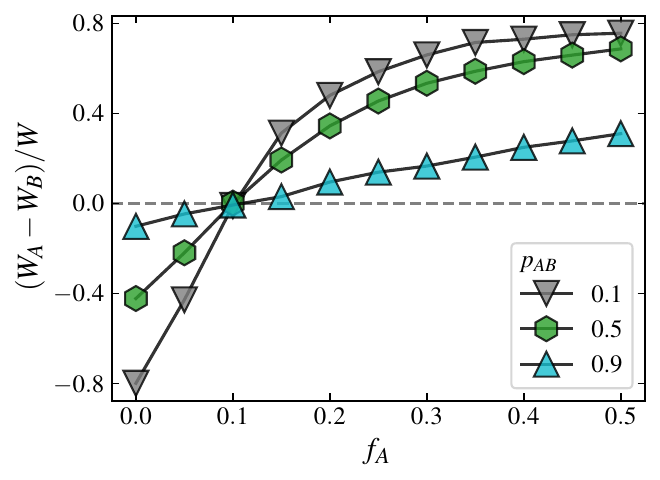}
    \caption{Wealth transferred from $B$ to $A$ normalized by the total available wealth for different as function of $f_A$. The other social protection factors, $f = f_B = 0.1$, remain fixed. The different $p_{AB}$ are indicated in figure. Lines are guides to the eyes.} 
    \label{fig:f5w}
\end{figure}

The impact of $f_A$ on the wealth transferred from $B$ to $A$ as function of $f_A$ is shown in figure~\ref{fig:f5w}. Social protection factors $f$ and $f_B$ were kept fixed in 0.1 and three values of $p_{AB}$ were considered. One can see that $W_A = W_B$ for $f_A = f_B = 0.1$ for all $p_{AB}$ we have used. The difference of wealth held by the groups is reduced with increasing $p_{AB}$. The exchanges between agents of distinct groups result in less chances to the wealth be redistributed within the groups accordingly their internal protection factors and the inter-group protection becomes more relevant. At the limit of $p_{AB} = 1$ (not shown in figure~\ref{fig:f5w}) all the trades are between an agent of $A$ and one of $B$; they obey the same rule and no net wealth transfer is observed.

\begin{figure}[!htbp]
    \centering
    \includegraphics[width = 0.9\textwidth]{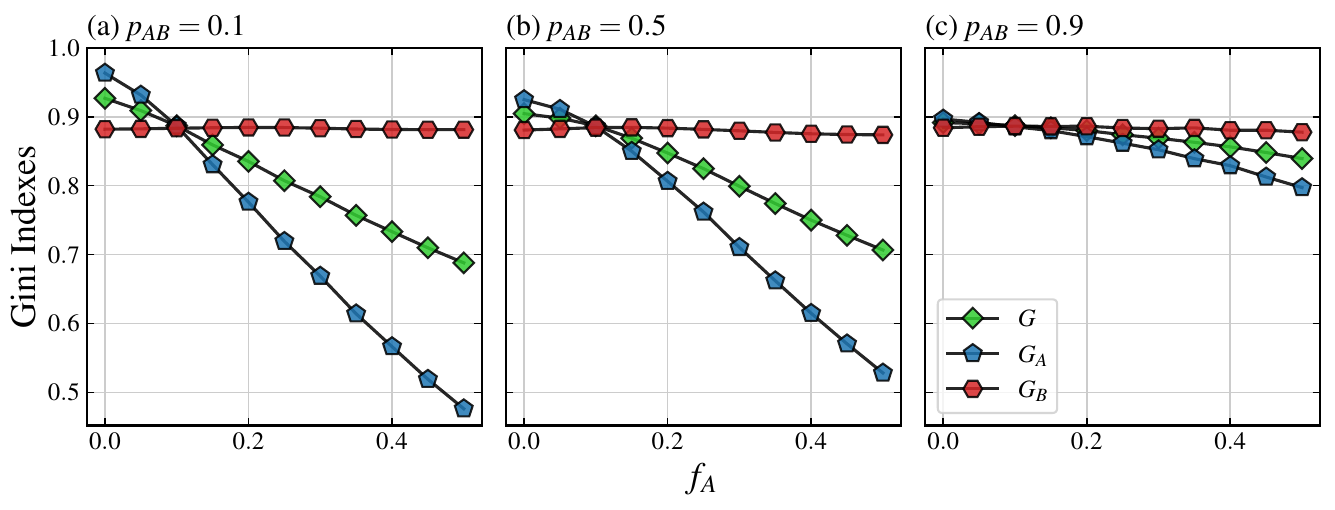}
    \caption{Gini indexes as functions of $f_A$ for $f = f_B = 0.1$. The inter-group exchange probabilities $p_{AB}$ are indicated in the figure. Solid lines are guides to the eyes.}
    \label{fig:f6g}
\end{figure}

Figure~\ref{fig:f6g} shows Gini indexes as functions of $f_A$ for $f = f_B = 0.1$. For $p_{AB} = 0.1$ our results for $G_A$ agree with the ones obtained in a society with indistinguishable agents \cite{iglesias}. The inequality of the complete system also decreases with $f_A$, mainly because group $A$ reducing inequality for the three $p_{AB}$ considered. Gini index of $B$ virtually does not change with the modifications on $f_A$. We see less variation of $G_A$ and $G$ when more inter-group trades take place. This is another feature related to the increase of the relevance of $f$ to the system, as discussed above. 

\begin{figure}[!htbp]
    \centering
    \includegraphics[width = 0.9\textwidth]{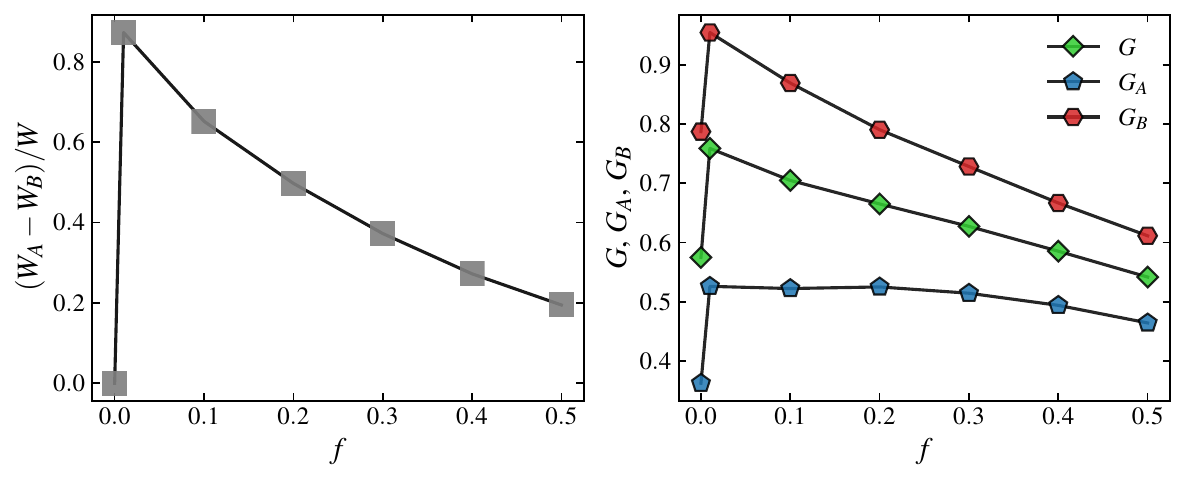}
    \caption{Left: Wealth transfer from group $B$ to $A$ as a function of the inter-group protection factor. Right: Gini indexes $G$, $G_A$ and $G_B$ as functions of the same quantity. The fixed parameters were $f_B = 0.1$ and $p_{AB} = 0.5$. Notice the qualitative similarities between wealth transfer and inequality curves.}
    \label{fig:f7f}
\end{figure}

The influence of the inter-group social protection factor on wealth transfer and inequality of groups was also investigated. Figure~\ref{fig:f7f} presents the wealth difference of $A$ and $B$ and the Gini indexes of both groups and the whole economy as functions of $f$ with $f_A = 0.5$, $f_B = 0.1$ and equally probable intra- and inter-group trades. One can notice that $(W_A - W_B)/W$ and the three inequality curves show similar behavior: a pronounced peak when $f = 0.01$ followed by a monotonic decrease. We already showed that non-zero inter-group protection to the poorest is a \emph{sine qua non} for the flow of wealth toward the most-protected group. 

Another unexpected feature we can identify in figure~\ref{fig:f7f} is that inequality also initially grows when a small protection is introduced to the inter-group trades. While the increase of internal protection factors drops the group inequality (and $G$ as a consequence), the growth of $f$ enhances the Gini indexes. For example, $G_A(f)$ is always larger than its value when $f = 0$. Group $B$ inequality starts with $G_B = 0.78$ with unprotected inter-group trades and quickly goes to 0.95 for $f = 0.01$. Only for $f \geq 0.3$ it gets lower than its initial value. 
 
\begin{figure}[!htbp]
    \centering
    \includegraphics[width = 0.8\textwidth]{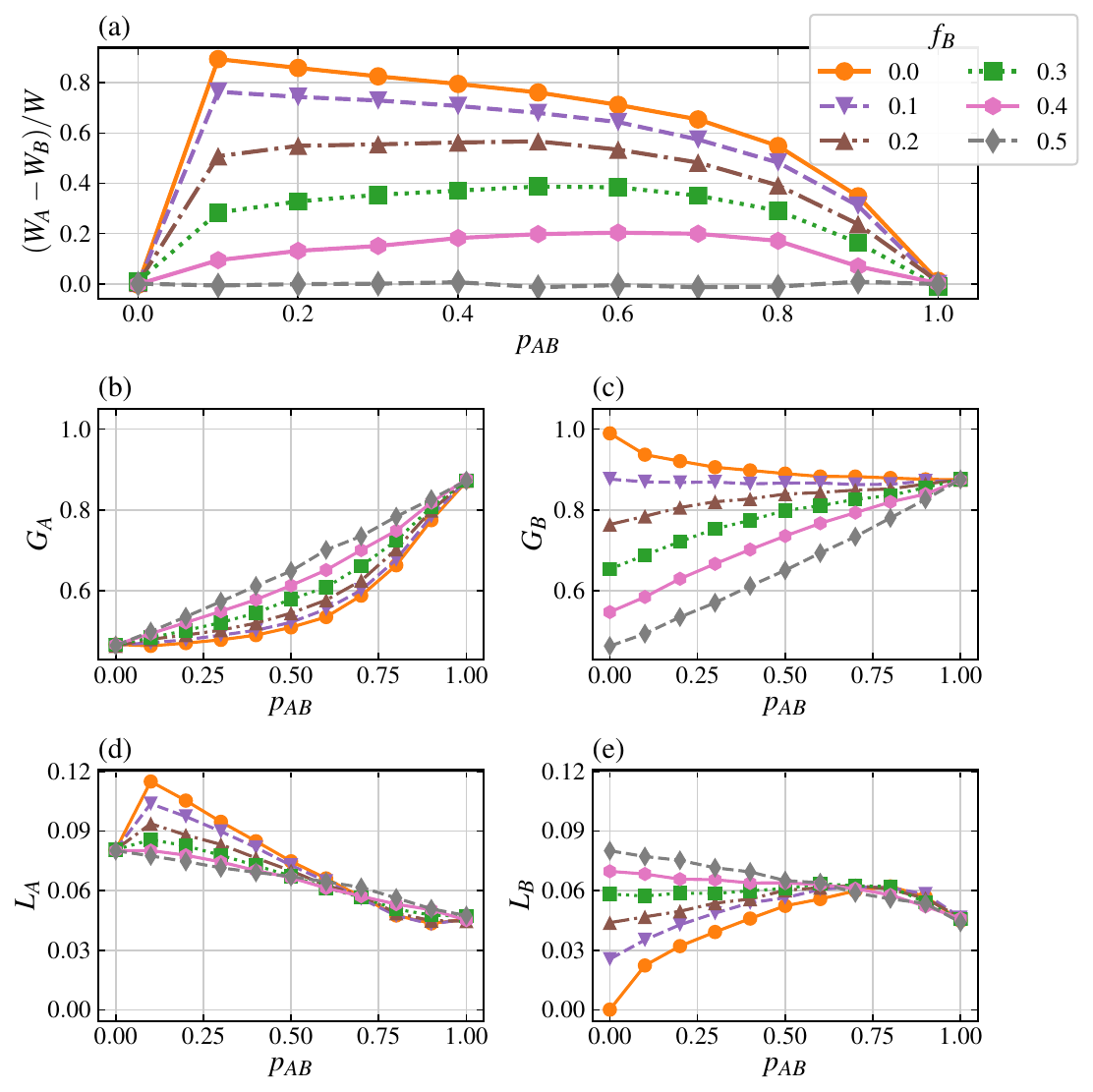}
    \caption{Impact of $p_{AB}$ on the transfer of wealth, Gini indexes, and liquidity of the groups. (a) Normalized wealth gap between the groups, (b) and (c) Gini indexes of groups $A$ and $B$, respectively. (d) $L_A$ and (e) $L_B$ as functions of $p_{AB}$. The $f_B$ for each curve is indicated in the figure. We have fixed $f = 0.1$ and $f_A = 0.5$. Lines are only guide to the eyes.}
    \label{fig:f8p}
\end{figure}

We also studied the influence of the inter-group exchange probability on the transfer of wealth between the groups. Gini indexes and economic mobility of the groups and their dependence on $p_{AB}$ were analyzed as well. We kept $f = 0.1$ and $f_A = 0.5$ fixed, $f_B$ was varied from 0 to 0.5. The results are compiled in figure~\ref{fig:f8p}. The analysis of the wealth transfer, figure~\ref{fig:f8p}\,(a) shows that $W_A = W_B$ for $p_{AB} = 0$ and 1 for all $f_B$ values. The first case corresponds to two isolated economies and $p_{AB} = 1$ indicates that every trade are between agents of different groups, respecting the general social protection $f$.

The total wealth of both groups is also equal when $f_A = f_B$, regardless the probability of inter-group trades. On the other hand, wealth flows from the less protected group to the other if the internal social protections are different. The amount transferred is larger as larger is this difference. For each $f_B$ the difference between groups' wealth presents a maximum at a given value of $p_{AB}$.  The horizontal position of the maxima is related with the difference between $f_A$ and $f_B$ in a way that larger discrepancies between these parameters present the largest gap in smaller $p_{AB}$. 

Gini indexes of the groups as functions of $p_{AB}$ are shown in figures~\ref{fig:f8p}\,(b) and (c). If $f_A = f_B = 0.5$ we see no differences between $G_A$ and $G_B$ and they increase almost linearly from 0.47 with no inter-group exchanges to $\approx 0.9$ when only inter-group exchanges take place. The shape of $G_A(p_{AB})$ became curvier as the $B$ protection reduces. The case of $G_B$ is more straightforward, if only intra-group trades are allowed $G_B(f_B) \approx 1 - f_B$ and the growth of $p_{AB}$ brings $G_B$ to 0.9 if only pairs of agents of different groups are chosen.

The last two panels of figure~\ref{fig:f8p} show the liquidity of the groups and their modifications with the probability of inter-group exchanges. We observe a peak in $L_A$ for $p_{AB} = 0.1$ and $f_B < 0.5$. This feature is a result of the transfer of wealth from $B$ to $A$. Since the group $A$ has more wealth to redistribute and it presents large social protection, its agents are able to exchange larger amounts of wealth between them at each MCS. The other group presents a more complex behavior of its liquidity. For $p_{AB} \geq 0.7$ the values of $L_B$ is almost independent of $f_B$. However, very different curves are seen when the inter-group exchange probability is smaller than 0.7. Depending on the $B$ social protection, it can be favorable to increase the relative number of exchanges with agents of $A$ within the interval $p_{AB} = [0, 0.6]$. The exceptions are $f = 0.3$ (where no significant changes in $L_B$ are noticed) and $f_B \geq 0.4$, which decreases with the increment of $p_{AB}$. 

Our findings indicate that it is possible to tune the best conditions for the unprotected group improve its liquidity and reduce inequality without loosing a considerable part of wealth.

\section{Summary and Conclusion}

The complexity of societies leads to their division into classes. Not only economical-related features are responsible for classify people into groups. These divisions impact on the wealth distribution and, thus, on the accumulation of wealth by one group at the expense of the others.  This work aims to investigate the microscopic mechanisms that generate inequalities and wealth transfer between two groups in an \emph{a priori} stratified society. 
The model considers two groups, $A$ and $B$, whose agents can interact by exchanging part of their wealth. Each group has an internal social protection that favors the poorest agent, which can be understood as a collective behavior that tends to maintain or advance the status of the group. Inter-group exchanges happen with probability $p_{AB}$ and we defined the protection of these trades as $f$.

The most-protected group is more equal and holds the biggest slice of the available wealth when compared with the other group, but only if the inter-group trades are regulated with $f > 0$. We found that there are two ways to avoid the transfer of wealth between the interacting groups, (i) by equalizing the protection factors, $f_A = f_B$; or (ii) if no protection acts over the inter-group exchanges. The first case also result in the same Gini index of both groups. The case of no inter-group protection maintain the different Gini index in a way that the less protected group is more unequal and promotes the condensation if there is no protection to the poorest agents.

We see that the group with less protection ($B$) can influence negatively the inequality of $A$ by increasing its protection. At the same time, it reduces the amount of wealth transferred. On the other hand, maintaining $B$ with low protection and changing $f_A$ does not impact significantly on the Gini of $B$. One can be conclude that the most vulnerable group depends only on its social protection to become more equal. On both cases, the whole economy benefits from the increase of one group's protection. 

Inter-group social protection also interferes on the economy of the groups. While $f = 0$ avoid the movement of wealth between groups, $f = 0.01$ causes the accumulation of resources in the protected group, which is accompanied by increases in $G$, $G_A$, and $G_B$. For larger values of $f$ these indicators tend to reduce. In a certain way, our results show that unprotected inter-group exchanges are better than the ones with small protection in terms of wealth transfer, inequality and economic mobility. We analyzed the effect of $p_{AB}$ on these measures.  It is favorable to the most-protected group trade with the other with small probability ($p_{AB} = 0.1$) since it increases the amount of wealth held by the group and, thus, the liquidity without change significantly its inequality. As the relative number of inter-group exchanges rises we notice that groups differences in Gini indexes, mobility and available wealth diminishes as a consequence of the importance of $f$ in spite of the intra-group protections.

The results of our simulations were compared with income distribution of Brazil divided by race, being White and Afro-Brazilian communities the two groups the model works with. The qualitative agreement shows that the model can be applicable to this case as well as to the merchandising between two countries.

\section*{Acknowledgements}

\ldots

\end{document}